\documentclass[12pt]{article}
\usepackage[dvips]{graphicx}
\usepackage{bm}
\usepackage{color}

\begin{document}
\title{Long-term impact risk for (101955) 1999 RQ$_{36}$}
\date{April 29, 2009}

\author{
Andrea Milani$^1$, Steven R. Chesley$^2$, Maria Eugenia Sansaturio$^3$, \\ 
Fabrizio Bernardi$^{4,1}$, Giovanni B. Valsecchi$^4$, Oscar Arratia$^3$ \\
\\
$^1$Department of Mathematics, University of Pisa \\
Largo Pontecorvo 5, 56127 Pisa, Italy \\
e-mail: milani@dm.unipi.it \\
$^2$Jet Propulsion Laboratory, Calif. Inst. of Tech.\\ 
Pasadena, CA 91109 \\
$^3$E.T.S. de Ingenieros Industriales, University of Valladolid\\
Paseo del Cauce s/n, 47011 Valladolid, Spain\\
$^4$IASF-Roma, INAF \\
via Fosso del Cavaliere 100, 00133 Roma, Italy
}

\maketitle

\begin{abstract}

The potentially hazardous asteroid (101955) 1999 RQ$_{36}$ has a
possibility of colliding with the Earth in the latter half of the 22nd
century, well beyond the traditional 100-year time horizon for routine
impact monitoring. The probabilities accumulate to a total impact
probability of approximately $10^{-3}$, with a pair of closely related
routes to impact in 2182 comprising more than half of the total. The
analysis of impact possibilities so far in the future is strongly
dependent on the action of the Yarkovsky effect, which raises new
challenges in the careful assessment of longer term impact
hazards. 

Even for asteroids with very precisely determined orbits, a future
close approach to Earth can scatter the possible trajectories to the
point that the problem becomes like that of a newly discovered
asteroid with a weakly determined orbit. If the scattering takes place
late enough so that the target plane uncertainty is dominated by
Yarkovsky accelerations then the thermal properties of the asteroid,
which are typically unknown, play a major role in the impact
assessment. In contrast, if the strong planetary interaction takes
place sooner, while the Yarkovsky dispersion is still relatively small
compared to that derived from the measurements, then precise modeling
of the nongravitational acceleration may be unnecessary. 

\end{abstract}

{\bf Keywords}: Asteroids, Asteroid dynamics, Impact processes, Orbit
determination.

{\bf Acronyms:}

IP: Impact Probability \cite{clomon2}

LOV: Line Of Variations \cite{multsol}

MC: Monte Carlo \cite{chodas}

TP: Target Plane \cite{analytic}

VA: Virtual Asteroid \cite{virimp}

VI: Virtual Impactor \cite{virimp}

\section{Impact Monitoring over centuries}
\label{s:intro}

As the completion of the {\it Spaceguard Survey} \cite{spaceguard}
approaches, we need to start thinking how the task of Impact
Monitoring \cite{clomon2} will change in the near future and
how we will face the next challenges. When almost all the Near Earth
Asteroids (NEAs) of a given size range have been discovered and found
not to be hazardous in the next 80--100 years, the problem of what
could happen over longer time scales remains.  Thus the need arises
for extending the Impact Monitoring work deeper into the future. This
also implies greater consideration of the sources of orbit propagation
uncertainty, for instance asteroid-asteroid perturbations or
nongravitational perturbations. For both of these, there are model
parameters that have to be solved in addition to the initial
conditions and that contribute to the prediction uncertainty.

For a well-observed asteroid, taking into account the Yarkovsky effect
and propagating for a long time span, the situation becomes in some
sense similar to that of a newly discovered asteroid, for which the
effects of chaotic motion and the resulting predictability horizon
increase the difficulty of detecting possible impacts.

When a NEA has a very well determined orbit, follow-up astrometric
observations may not be necessary 
{\it now}: 
if the current ephemeris error
is smaller than the observational errors, the orbit is not
significantly improved. If the main sources of prediction uncertainty,
over the time span to the possible impact, are the nongravitational
effects, then physical observations and theoretical modeling are the
most important contribution to solve the problem.

An example of the situation above has already occurred with
the identification of a very long term possibility of impact for the
asteroid (29075) 1950DA \cite{1950DA}. It has been observed
optically for more than 50 years and with radar in 2001; its semimajor
axis has a formal uncertainty (post-fit RMS)
of $\simeq 100$ m, the ephemeris
uncertainty is $<0''.05$, thus it is currently impossible to improve
the orbit by optical astrometry. For the possible impact
in 2880 the main source of uncertainty is the dynamical model,
especially the Yarkovsky effect. 

This paper is organized as follows: in
Section~\ref{s:rq36} we present the case of asteroid (101955) 1999
RQ$_{36}$ and explain why it is especially interesting for long term
impact monitoring. In Section~\ref{s:yarko} we discuss a model for the
Yarkovsky effect on this asteroid and its
uncertainties. Section~\ref{s:montecarlo} presents our Monte Carlo
test to detect possible impacts taking into account the main
non-gravitational effects. In Section~\ref{s:LOVsampling} we compare
with the results which can be obtained with the current operational
impact monitoring methods, in particular explaining why impacts in the
year 2182 have a comparatively high
probability. Section~\ref{s:interpretation} provides an
interpretation of the dynamical behaviour of (101955) as a sequence
of phases dominated by different sources of orbit
uncertainty. Section~\ref{s:next} discusses the possibility and
difficulty of improving the orbit determination at the next
apparitions, thus significantly changing the estimates on impact
probability. Section~\ref{s:deflection} discusses how difficult is to
deflect this asteroid, in case the possibility of impact was confirmed
by later observations, and which is the most appropriate time to do
this. The last Section draws our conclusions, including the one that
the case of (101955) is not exceptional, in that it is already
possible to find another example for which possible collisions in the
22nd century can be obtained.

\section{Asteroid (101955) 1999 RQ$_{36}$}
\label{s:rq36}

We need to select a good example as a test case to develop the
analytical tools and the know-how for very long term Impact
Monitoring. 

The Apollo
asteroid (101955) 1999 RQ$_{36}$, having been observed
astrometrically over almost 7 years and with radar at two separate
apparitions, has the lowest formal uncertainty in semimajor axis of
any asteroid\footnote{This minimum uncertainty refers to the initial
conditions at the epoch corresponding to the average of the
observation times, that is 8 June 2001. For these data, see the NEODyS
online system {\tt http://newton.dm.unipi.it/neodys}}: $5$ m.
However, this uncertainty is indeed formal, in that it ignores the
dynamical model uncertainty.  As we will show later, a reasonable
estimate of the Yarkovsky effect on the semimajor axis of 1999 RQ$_{36}$ is
around $200$ m/yr. Including a Yarkovsky effect model in the orbit
determination can change the nominal solution for the initial
conditions by an amount more than an order of magnitude larger than
the formal uncertainty.

\begin{figure}[htb]
\centerline{\includegraphics[width=12cm]{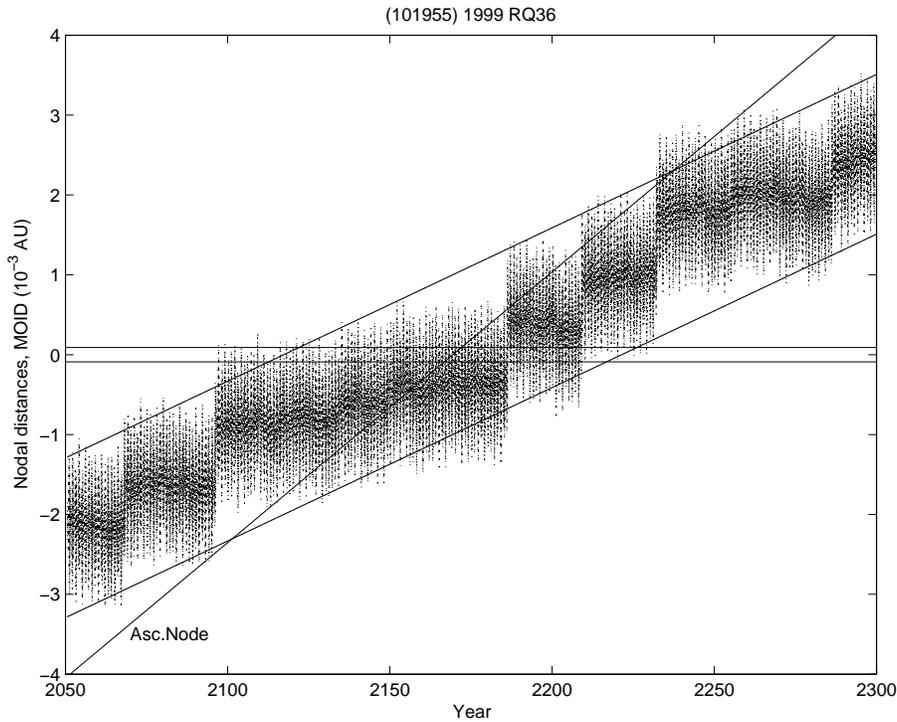}}
\caption{The time evolution of the MOID for (101955) 1999
RQ$_{36}$. The MOID is often less than the effective radius of the Earth
(horizontal lines), starting a little before 2100 and until
about 2230.}
\label{fig:moid}
\end{figure}

The reason why this asteroid is relevant for long term Impact
Monitoring is that its Minimum Orbit Intersection Distance (MOID)
\cite{moid2} is currently low (about one lunar distance)
and decreases to near zero at the end of next century. This is shown
in Fig.~\ref{fig:moid}, where---to stress the secular evolution---the
MOID is represented with sign according to whether the asteroid is
outside (positive) or inside (negative) the Earth's orbit at the node
crossing.
This convention on the sign is as in \cite{analytic}, and is
the opposite with respect to \cite{moid_reg}. The Figure also includes a
line (marked ``Asc.Node'') representing the secular evolution of the
nodal distance and two parallel lines representing the secular
evolution of the signed MOID. 

Thus the secular evolution leads the MOID to cross zero around 2170,
with the short periodic perturbations allowing a MOID less than the
Earth impact parameter (equal in this case to $2.12$ physical Earth
radii) at various epochs between 2097 and 2229. This analysis applies
to the purely gravitational, nominal solution, 
for other possible orbits (compatible with
the observations) the MOID evolution is similar, that is a node
crossing occurs around the same time, although the epochs at which the
collisions are possible are different.  The node crossing and the
growth with time of the along-track position uncertainty are the
parameters used in the criteria used to prioritize the Impact
Monitoring computations \cite[Sec. 2.2]{clomon2}. In other words this
asteroid, with currently a very well known orbit, becomes in the
22nd and early 23rd century like a newly discovered \emph{Potentially
Hazardous Asteroid} (PHA), with a strongly chaotic orbit. 

Considering the discussion above, we have run an Impact Monitoring
computation for (101955) 1999 RQ$_{36}$ in the conventional way 
(with only gravitational effects in the dynamical model) 
but for a time span
longer than usual, until the year 2240. The results were that there
are Virtual Impactors (VIs), that is, sets of initial conditions compatible with
the observational constraints leading to impacts, in the 2180-2200 time 
span and with a total Impact Probability (IP) $\simeq 3\times 10^{-4}$.

However, for the reasons discussed in the previous section, this is
not an appropriate way to assess the possibility of an
impact and is even poorer for computing the IP. We need to take into
account the probabilistic distribution of the parameters appearing in
a Yarkovsky model.

\section{Yarkovsky effect model for 1999 RQ$_{36}$}
\label{s:yarko}

The main nongravitational perturbation over a time span of centuries
is due to the Yarkovsky effect, which is by definition a secular
effect and results from the way the asteroid rotation affects the
surface temperature distribution and therefore the anisotropic thermal
re-emission \cite{david00}. The primary manifestation of the Yarkovsky
effect is through a steady drift in semimajor axis:
\[
\frac{da}{dt} \propto\frac{\cos \gamma}{\rho R},
\] 
where $\rho$ is the asteroid's bulk density, $R$ is the asteroid
effective radius and $\gamma$ is the angle between the orbital and
rotational angular momentum vectors, i.e., the obliquity of the
asteroid's equatorial plane with respect to its orbital plane. There
is also a complex dependency on the spin rate and
surface thermal conductivity $K$,
but the main parameter upon which a Yarkovsky model depends is
$\cos \gamma$. Thus the sign of the drift in semimajor axis changes
when the sense of rotation changes, and the problem is that optical
ground-based observations only weakly discriminate the sense of
rotation.

Asteroid (101955) 1999 RQ$_{36}$ has been carefully observed during
the 2005 encounter with Earth, with optical photometry, radar range
and range rate observations. 
Table~\ref{t:poles} lists the assumed model
parameters affecting the Yarkovsky effect.
The rotation period has been well
determined through light-curve analysis (Carl
Hergenrother, private communication). For the pole orientation, the
analysis of the radar echoes leads to the 
retrograde solution listed in Table~\ref{t:poles} 
(\cite{nolan}, Nolan, private communication) and 
the mean radius has also been estimated at $280$ m.  

\begin{table}[htb]
\caption{Pole solution and other parameters relevant for the Yarkovsky 
effect, as assumed in our model for 1999 RQ$_{36}$.}
\begin{center}
\label{t:poles}
\begin{tabular}{lcc}
\hline
Parameter & Value & Units \\
\hline
Pole RA &$102.37$& deg\\
Pole DEC&$-52.85$& deg\\
Obliquity $\gamma$&$169.9$& deg\\
Radius $R$ &$280$& m\\
Albedo &$0.05$ &\\
Rotation period& $4.2903$& h\\
Bulk density $\rho$& $1,500$& kg/m$^3$\\
Surface layer density& $1,200$&kg/m$^3$\\ 
Thermal conductivity $K$ & $0.01$&W/m/K\\
\hline
\hline
\end{tabular}
\end{center}
\end{table}

From this model it is possible to deduce an a priori estimate of the
secular change in semimajor axis $da/dt$, taking Nolan's spin state as
accurate enough to contribute little to the uncertainty. With that,
the main uncertainties come from density $\rho$, radius $R$ , and
thermal conductivity $K$.

First, $da/dt$ varies inversely with $\rho\,R$. Taking the formal
uncertainty in $\rho$ to be $30\%$ and in $R$ to be $15\%$ (these are
merely educated guesses) we find that this leads to a $34\%$
uncertainty on $da/dt$. A $20\%$ uncertainty in $da/dt$
due to variations in $K$ can be
assumed, because $da/dt$ is only weakly sensitive to $K$; this gives
$40\%$ uncertainty on $da/dt$, altogether. The obliquity of
$170^\circ$ obtained from Nolan's pole has some uncertainty, but
because the relevant parameter is the cosine of the obliquity this has
a modest effect on the uncertainty of the diurnal Yarkovsky
contribution. With a nominal $K=10^{-2}$ W/m/K, and following the
techniques of \cite{david00}, we obtain
\begin{equation}
\frac{da}{dt}=-12.5 \pm 5 \times  10^{-4}\; \mathrm{AU/Myr}\ .
\label{eq:dadt}
\end{equation}

\begin{figure}[h]
\centerline{\includegraphics[width=10cm]{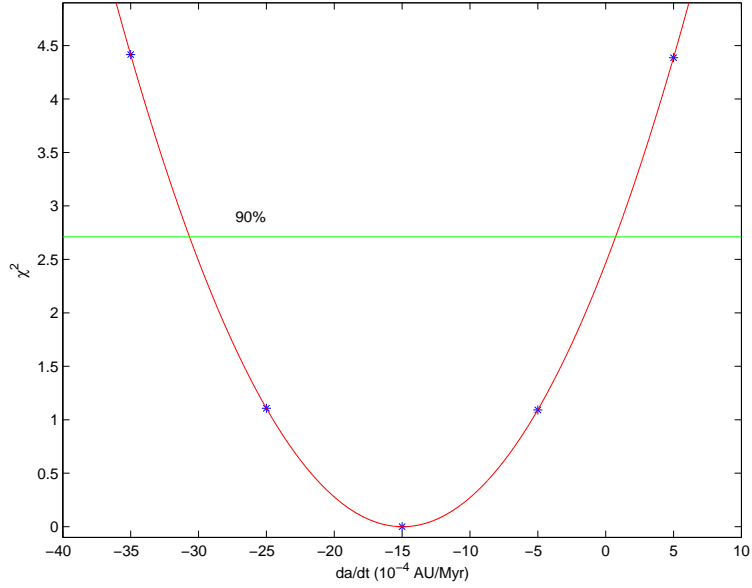}}
\caption{The $\chi^2$ of the least squares fit to the value of
$da/dt$. The quadratic approximation for the $\chi^2$ is obtained by a
parabola fit to only 5 data points (stars). The horizontal line
corresponds to a $90\%$ confidence level for this fit.}
\label{fig:yarkofit}
\end{figure}

As an alternative, a plausible range of values for $da/dt$ could be
estimated by least squares fit to the astrometric data, including both
optical and radar observations. Figure~\ref{fig:yarkofit} shows a
simple method to perform such an estimate, by using 5 separate orbit
determinations\footnote{ This fit has used the free software OrbFit,
version 3.5.2. OrbFit is available  at
\tt{http://adams.dm.unipi.it/orbfit/}}, each one with a different (but constant) value of
$da/dt$. The signal is contained in the different values of the fit
RMS, which has the value $0.524$ assuming no Yarkovsky effect and
$0.520$ at the minimum. The best fit value is
\[
\frac{da}{dt}=-15.0 \pm 9.5 \times  10^{-4}\; \mathrm{AU/Myr}\, ,
\]
which means that for now we have only a weak detection of the
Yarkovsky effect. A similar result has been reported by
\cite{chesley08}: the two results are actually identical
if the different covariance normalization is taken into account. In the
computation of Fig.~\ref{fig:yarkofit} a simple error model, with
optical observations weighted at 1 arcsec, was used. In conclusion the
a priori estimate is superior to the ones based on the orbital fit,
thus we adopt Eq.~(\ref{eq:dadt}) as a plausible range.

\section{Monte Carlo tests}
\label{s:montecarlo}

To assess the possibility of impacts taking into account the Yarkovsky
effect with its full uncertainty, we have performed \emph{Monte Carlo}
(MC) simulations with the method described in
\cite{chodas}. The MC samples are drawn from a
7-dimensional space of initial conditions, including 6 orbital
elements and $da/dt$, based on a normal distribution derived from the
a posteriori fit covariance, as obtained from the available
observational data and the a priori constraint given by
Eq.~(\ref{eq:dadt}).

\begin{figure}[!h]
\centerline{\includegraphics[width=12cm]{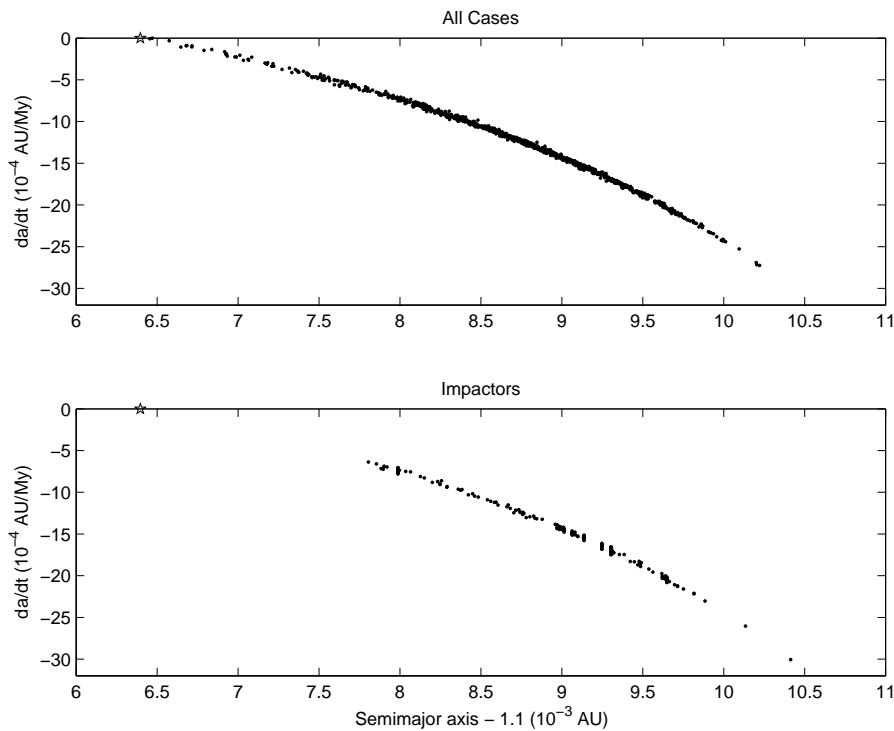}}
\caption{Above: a swarm of Virtual Asteroids (VA), among the $500,000$
used in our main Monte Carlo test, plotted in the plane $da/dt$
vs. $a-1.1$ AU at epoch January 1, 2081. Below: the orbits leading to an
impact. The star represents the purely gravitational solution. }
\label{fig:after80_adadt}
\end{figure}

Each different MC sample follows its own dynamical route, although the
individual sample orbits remain close together for a long time; only after the
encounters with the Earth in 2060 and 2080 do they become widely scattered.
In Fig.~\ref{fig:after80_adadt} the orbits are plotted
shortly after the 2080 encounter, by which time the values for the
semimajor axis have been scattered over a range of $0.005$ AU.

\begin{figure}[!h]
\centerline{\includegraphics[width=12cm]{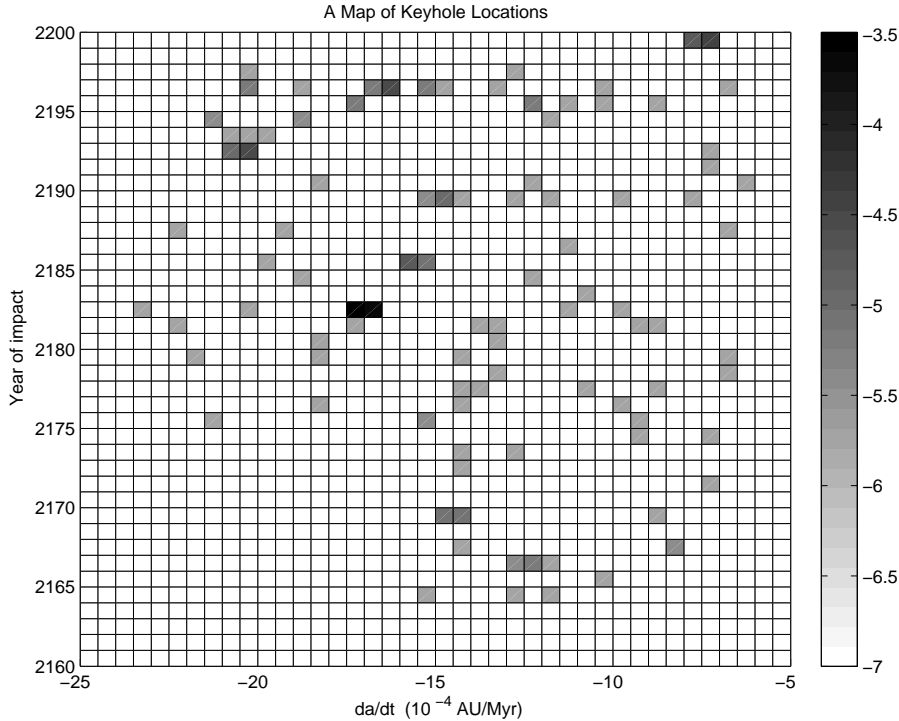}}
\caption{The distribution of the impactors found in the MC run as a
function of the secular drift in semimajor axis due to Yarkovsky and
the impact year. The shading is according to $\log_{10}$
of the probability of impact obtained for each cell.}
\label{fig:adot_yr_hist_2}
\end{figure}

Out of a sample of $500,000$ initial conditions, which were propagated
up to the year 2200, we have found 461 cases of impact. This implies
that the overall IP is $\simeq 9.2\times 10^{-4}$, more than what was
obtained by neglecting the Yarkovsky effect
altogether. Figure~\ref{fig:adot_yr_hist_2} shows the distribution as
a function of the Yarkovsky $da/dt$ and of the year of impact.
Figure~\ref{fig:adot_yr_hist_2} depicts the location and impact
probability of the various distinct dynamical routes to impact that
are revealed by the Monte Carlo analysis.

Looking for the most significant VIs, we have found that
272 of the $461$ impacting Virtual Asteroids
(VA) have an impact in the year 2182. Of these, 268 have similar
dynamical histories: they appear in Fig.~\ref{fig:after80_adadt} as a
narrow strip with $1.1093021<a<1.1093076$ AU. Thus there is a single
dynamical route leading to an impact in 2182 with an Impact
Probability estimated, by the Monte Carlo method, at $\simeq 5.4\times
10^{-4}$.

Table~\ref{t:rq36_risk} contains additional details for the cases
of the VIs with highest impact probabilities. Note that the values
listed in the Table have not been computed with an automatic, 
well-documented algorithm, but by ad hoc manipulations. 
For example, the values of
the IP are obtained from the number of impacting VA in the MC run
divided by the total number in the sample, the value of the stretching
has been deduced from the IP by a simple 1-dimensional formula,
similar to the ones of \cite[Section 4.3]{pssimp}:
\[
S= \frac{C\times \exp(-\sigma_{LOV}^2/2)}{IP\times \sqrt{2\,\pi}} \ ,
\]
where $C$ is the chord length corresponding to the crossing of a
sphere of radius $2.12\; R_\oplus$ at a distance from the center equal
to the minimum occurring for each VI. The value of $\sigma_{LOV}$ has
been computed on the basis of the value of $da/dt$.

\begin{table}[htb]
\caption{Most Significant VIs for (101955) 1999 RQ$_{36}$ through Year 2200}
\begin{center}
\label{t:rq36_risk}
\begin{tabular}{cccccc}
\hline
$da/dt$            & Year & $\sigma_{LOV}$ & Stretch      & IP & Palermo \\
(10$^{-4}$ AU/Myr) &      &                & ($R_\oplus$) &    & Scale   \\
\hline
$-14.60$ & 2169/9/24.72 & $-0.42$ & $9.63\times10^4$ & $1.60\times10^{-5}$ & $-2.73$ \\
$-17.10$ & 2182/9/24.93 & $-0.92$ & $3.91\times10^3$ & $2.60\times10^{-4}$ & $-1.55$ \\
$-17.10$ & 2182/9/24.93 & $-0.92$ & $3.20\times10^3$ & $2.76\times10^{-4}$ & $-1.52$ \\
$-15.43$ & 2185/9/24.60 & $-0.59$ & $5.45\times10^4$ & $2.60\times10^{-5}$ & $-2.56$ \\
$-15.02$ & 2189/9/24.62 & $-0.50$ & $8.07\times10^4$ & $1.60\times10^{-5}$ & $-2.78$ \\
$-20.44$ & 2192/9/24.35 & $-1.59$ & $1.08\times10^4$ & $4.40\times10^{-5}$ & $-2.34$ \\
$-16.27$ & 2195/9/24.34 & $-0.75$ & $6.33\times10^4$ & $2.00\times10^{-5}$ & $-2.70$ \\
$-7.51$  & 2199/9/25.05 & $+1.00$ & $1.89\times10^4$ & $5.40\times10^{-5}$ & $-2.27$ \\
\hline
\end{tabular}
\end{center}
\end{table}

As reflected in Table~\ref{t:rq36_risk} and
Fig.~\ref{fig:adot_yr_hist_2}, there are many more VIs for different
years, but the year 2182 dominates the overall IP, so much that we
wish to identify the apparently special dynamical behavior that would
cause such a large region in the initial conditions space to impact.

\newpage

\section{Line of Variations Impact Monitoring}
\label{s:LOVsampling}

The standard method used for impact monitoring is the \emph{Line Of
Variation} (LOV) search \cite{clomon2}, based on the computation of
multiple initial orbits sampling the LOV, which is essentially a line
of weakness in the initial conditions space. The LOV definition
depends upon the choice of coordinates \cite{multsol}; in this case we
have used Keplerian elements and a scaling such that the LOV is
essentially along the $a$ axis, as it is appropriate when
the time span is so long that the along track runoff is the dominant
uncertainty.

We have used the CLOMON2 version of the impact monitoring algorithm,
in use at the Universities of Pisa and Valladolid, which is also
available in the free software \emph{OrbFit}; the SENTRY version, in use at
JPL, is based on similar principles, although implemented differently
\cite{clomon2}.  These software robots have been developed to monitor
possible impacts in the medium term, say 10 to 100 years from now, and
thus do not currently have the capability of fully taking into account
Yarkovsky and other nongravitational perturbations. For example, we
cannot yet sample a LOV in 7-dimensional space and use it to find VIs
in that space. We can include different forms of Yarkovsky effect in
the dynamical model, including the assumption of an acceleration in
the transversal direction, with an $1/r^2$ dependence upon the
heliocentric distance $r$, which results in a secular $da/dt$. The
constant value of $da/dt$ can be set for each run, thus we can explore
a 6-dimensional section in the 7-dimensional space of the MC runs.

\begin{figure}[!h]
\centerline{\includegraphics[width=12cm]{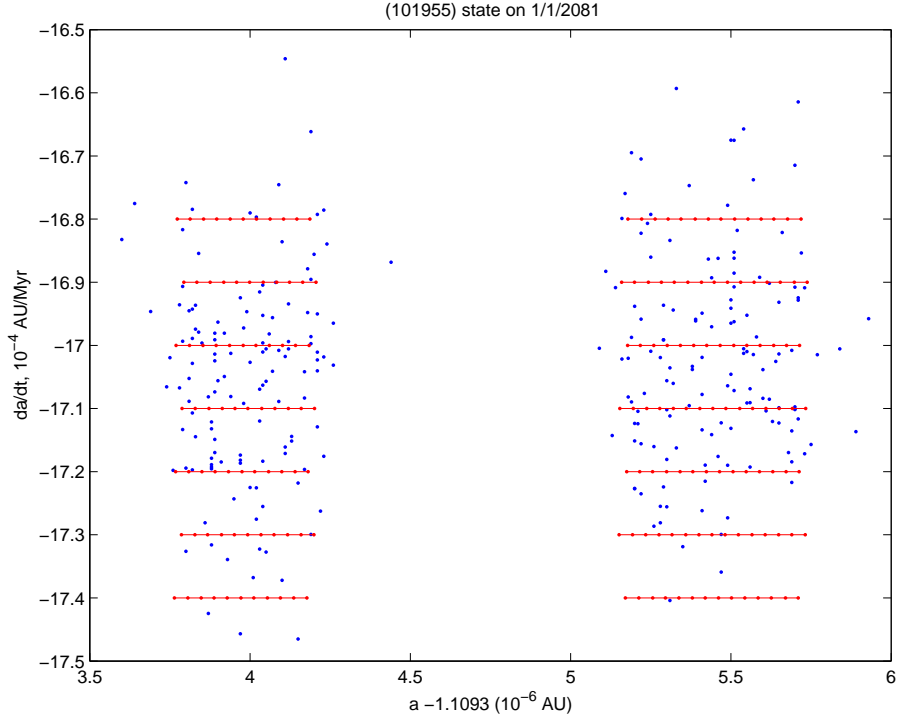}}
\caption{The Virtual Asteroids impacting in 2182, in the same plane as
  Fig.~\ref{fig:after80_adadt}, with $da/dt$ and semimajor
  axis at epoch January 1, 2081, that is after the 2080 close
  approach. The MC impactors are represented as isolated points, the
  consecutive LOV sample orbits which impact are represented as
  connected strings.}
\label{fig:after80VIs}
\end{figure}

We have performed 7 runs of CLOMON2, involving 2401 VAs
each, with values of $da/dt$ ranging from $-17.4$ to $-16.8$ with
step $0.1$ in units of $10^{-4}$ AU/My. This samples
the range of $da/dt$ where most of the MC impacts in 2182
are: Figure~\ref{fig:after80VIs} shows the MC sample orbits with
impact in 2182 and the LOV sample orbits with impact in 2182. The
latter are joined by segments whenever they belong to consecutive
sample points on the LOV to stress that our impact monitoring uses
geometric sampling, a version of \emph{manifold dynamics}, in which we
can interpolate smoothly between sample points.

The CLOMON2 algorithm can indeed interpolate and find, for each of the
14 segments, the point corresponding to the closest approach to the
center of the Earth and compute the estimated IP: the values obtained
in this way for 2182 the impacts range between $1.6\times 10^{-3}$ and
$2.1\times 10^{-2}$.  These are not correct estimates of the impact
probability under the present level of information about the orbit,
because they are conditional probabilities for a given, fixed value of
$da/dt$; still these values give an indication of what could happen if
the value of $da/dt$ was well known.

The most striking feature of Fig.~\ref{fig:after80VIs} is the
presence of two VIs, that is two separate components, each connected,
in the region of initial conditions (in the 7-dimensional space)
leading to an impact in 2182. This could be figured out by using the
MC points, but is demonstrated rigorously by using the LOV
sampling. The VAs consecutive as LOV sample points leading to a 2182 impact
are connected by a line segment; however, between the segments on the
left of the figure and the ones on the right, there are Virtual Asteroids
on each LOV not leading to an impact.

\begin{figure}[!h]
\centerline{\includegraphics[width=12cm]{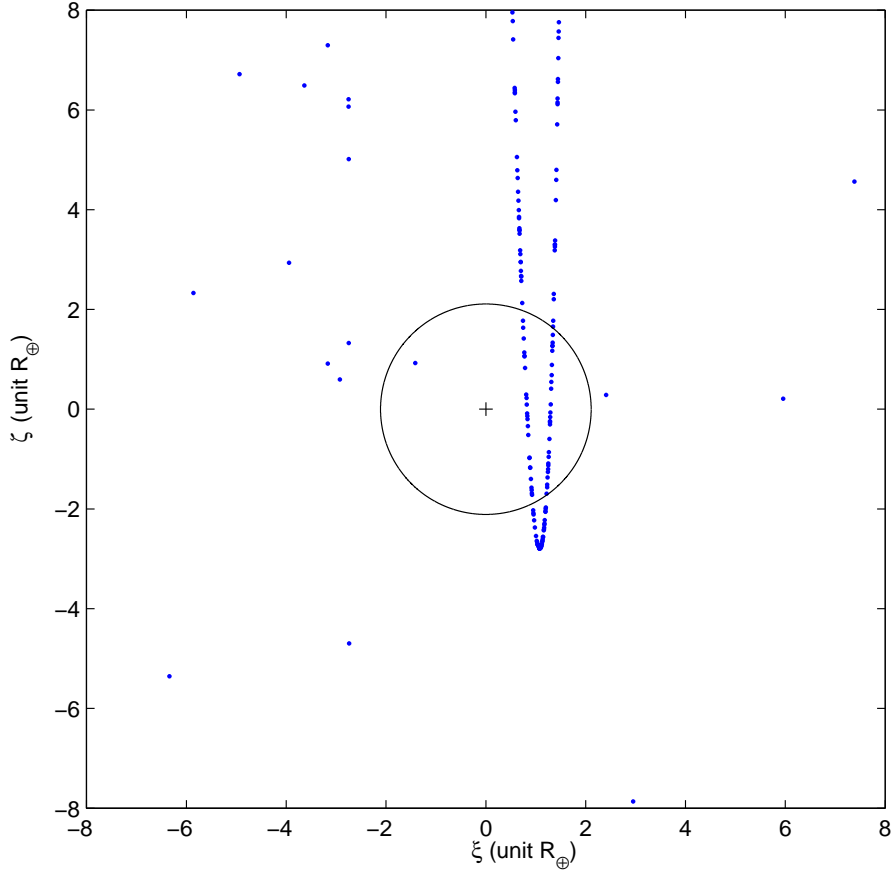}}
\caption{The Target Plane traces of the MC sample orbits which have a
very close approach in 2182; the coordinates are as in
\cite{analytic}, units are Earth radii.}
\label{fig:bpl2182}
\end{figure}

The explanation can be found by looking at the 2182 \emph{Target
Plane}\footnote{Hereafter we use TP coordinates defined as in
\cite{analytic}.}  (TP) trace of the MC and LOV computations. In
Fig.~\ref{fig:bpl2182} we show the MC points on the 2182 TP, including
the impactors and also the near-miss points. Note that, although the
MC points have been selected at random, according to a probability
density, in the 7-dimensional space, still most of the impactors and
near-miss points are organized in a very narrow strip entering the
figure from the top, crossing the Earth impact cross section, and
leaving the latter from the bottom. The strip then reverses,
crossing again the impact cross section and reaching back to the upper
boundary of the plot window. A similar figure done with the 7 LOV
sampling runs shows 7 strings doing the same as the MC strip: they are
so tightly packed that the human eye would not be able to separate
them at the scale of the plot.

The key point is that the very same reason why there are two separate
VIs is also the cause of the especially high IP associated to the two
VIs.  Indeed, if the LOV strings ``stop and come back'', this implies
that the derivative $d\zeta/d\sigma$ of the vertical coordinate in the
Figure with respect to the LOV parameter $\sigma$ 
has a zero
\footnote{Note that 
the $\sigma$ used here to parameterize the LOV is from CLOMON2 runs for 
fixed $da/dt$, and thus is not the same as that of Table~\ref{t:rq36_risk}.}, 
then
changes sign. 
This is called an \emph{interrupted return}
\cite{an10,clomon2,analytic}.  Since $d\zeta/d\sigma$ is a smooth
function, it has smaller values near the tip. This also affects the
\emph{stretching} $S$, which is the length of the derivative of the TP
point with respect to the LOV parameter $\sigma$. In most
points (except near the tip) the contribution of $|d\zeta/d\sigma|$
in $S$ is dominant.  However, $S$ is sharply reduced near the tip,
thus the TP points near the tip are closer to each other.
Because the tip where the strings turn back is very close to the
segments contained in the Earth cross section, more sample points
belong to the two VIs.

\begin{figure}[h]
\centerline{\includegraphics[width=10cm]{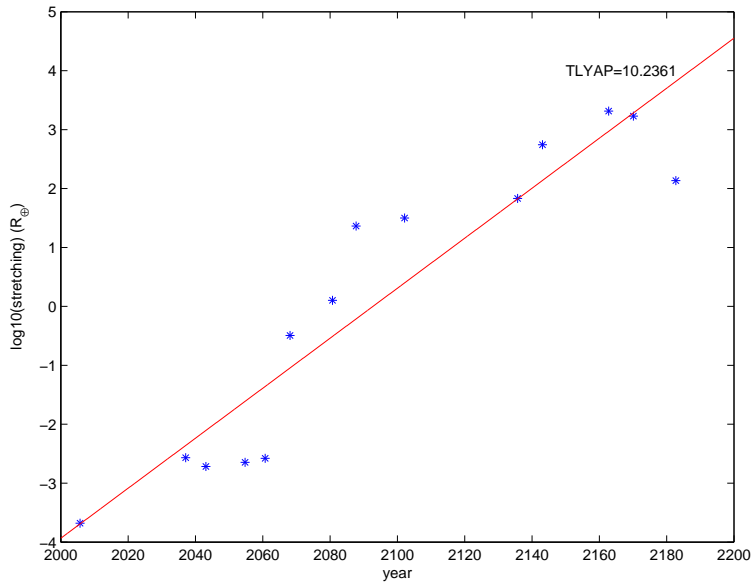}}
\caption{The stretching (in a $\log_{10}$ scale) generally increases
  at each close approach with Earth, but not at all of them. These
  data refer to the sample orbit of 1999 RQ$_{36}$ nearest to the
  center of the VI on the left in Fig.~\ref{fig:after80VIs}. The
  stretching on average grows by an order of magnitude in 23.6 years,
  thus by a factor $\exp(1)$ on average every $10.2$ years. }
\label{fig:lyap}
\end{figure}

To assess the importance of this phenomenon, we have plotted in
Fig.~\ref{fig:lyap} the behavior of $S$ at successive close approaches
with Earth.  The stretching generally increases, by a factor $\exp(1)$
on average every $10.2$ yr (this could be used as an estimate of the
so called \emph{Lyapounov time}). The Figure in fact shows that the
stretching increases by about 1 order of magnitude until 2060, then
the close approaches in 2060, 2068 and 2080 increase it by 4 orders of
magnitude, followed by moderate growth until 2170. After the close
approach of 2170 there is a decrease by a factor $\simeq 16$, which is
one way of explaining why the two largest VIs for 2182 have an IP an
order of magnitude larger than all of the others found in the MC run.

\section{Interpretation}
\label{s:interpretation}

To explain the very complex dynamics leading to especially large VIs,
we shall partition the time span from the initial conditions time in
2001 to the impact in 2182 in 4 phases, using the previous Figures,
Table~\ref{t:CAS} with the close approaches for one of
the 268 VAs with similar close approach sequence impacting 2182 and
the following order of magnitude argument: a Yarkovsky perturbation of
the order of $10^{-3}$ AU/My corresponds to $15$ km/100 yr.  The
discussion hereafter concerns the region of the LOV encompassing the
2182 VIs.

\begin{table}[htb]
\caption{Close Approach Summary for a VA impacting in 2182.}
\begin{center}
\label{t:CAS}
\begin{tabular}{lcccc}
\hline
 Date & CA Dist & Vinf & Stretch.  \\
 & AU & km/s & R$_\oplus$  \\  
\hline
2005 Sep 20.45  & $0.0331$ & $6.8$ & $2.30\times10^{-4}$\\
2037 Feb 11.56  & $0.0987$ & $7.5$ & $3.79\times10^{-3}$\\
2043 Feb 09.76  & $0.0966$ & $4.3$ & $1.66\times10^{-3}$\\
2054 Sep 30.04  & $0.0393$ & $5.1$ & $1.62\times10^{-3}$\\
2060 Sep 23.03  & $0.0050$ & $6.1$ & $2.71\times10^{-3}$\\
2068 Feb 15.14  & $0.0705$ & $5.7$ & $3.16\times10^{-1}$\\
2080 Sep 22.09  & $0.0149$ & $6.3$ & $1.38\times10^{+0}$\\
2087 Sep 29.83  & $0.0418$ & $5.0$ & $1.69\times10^{+1}$\\
2102 Feb 17.22  & $0.0779$ & $4.9$ & $2.38\times10^{+1}$\\
2135 Sep 27.82  & $0.0222$ & $5.5$ & $5.79\times10^{+1}$\\
2143 Feb 16.44  & $0.0776$ & $5.0$ & $4.41\times10^{+2}$\\
2162 Oct 19.05  & $0.0936$ & $4.4$ & $1.14\times10^{+3}$\\
2170 Feb 17.69  & $0.0927$ & $7.5$ & $2.63\times10^{+3}$\\
\hline
\hline
\end{tabular}
\end{center}
\end{table}

In phase 1 (2001--2060), the change in $a$ due to Yarkovsky is much
larger than the conditional uncertainty (for fixed $da/dt$) of the
semimajor axis: $RMS(a)\simeq 5$ m. Thus the solutions with different
values of $da/dt$ are well separated. Upon arrival on the Target Plane
of the 2060 encounter, the traces on the TP of the 7 VIs of our 7
CLOMON2 runs are still well separated, by $\simeq 80$ km, while the
segments of the LOV leading to the 2182 impact are only a few hundred
meters long.

In phase 2 (2060--2080), a sequence of 3 close approaches, in 2060
near the ascending node, in 2068 near the descending node, and in 2080
near the ascending node again, increases the stretching (as measured
at the next encounter in 2087) by a factor $\simeq 9,000$.

In phase 3 (2080--2162), the orbits compatible with the current
observational data are scattered widely and follow different dynamical
routes.  The spread in semimajor axis after the 2080 encounter can be
seen in Fig.~\ref{fig:after80_adadt}, top, for the whole set of VAs
used in the MC run, and is $\simeq 600,000$ km.  For the subset
leading to the main 2182 VIs, as shown in Fig.~\ref{fig:after80VIs},
the range of values in $a$ is $\simeq 90$ km wide for the larger VI
(on the right) and $\simeq 60$ km for the one on the left. Thus the
total effect of Yarkovsky on the semimajor axis in the time span
2081--2182 is less than the span in $a$ of the VIs, and quite
negligible with respect to the spread in 2081.  From there on, it is
the value of the semimajor axis that determines which dynamical route
will be followed, that is the sequence of close approaches to the
Earth, while the value of $da/dt$ will not have much influence, as it can only
displace slightly the position of the keyhole on exit from the 2080
encounter.

\begin{figure}[htb]
\centerline{\includegraphics[height=10cm]{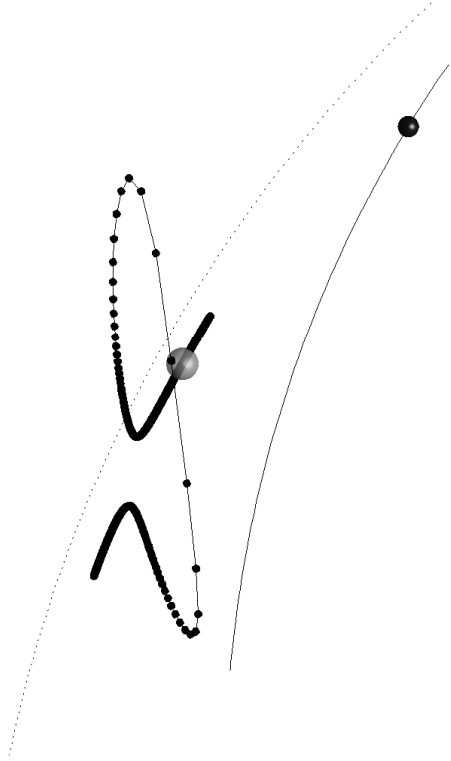}}
\caption{The position, projected on the ecliptic plane, of the VAs
forming a segment of the LOV 5 months after the September 2162 close
approach to the Earth (the dark sphere, not to scale). The transparent
sphere marks the segment of the LOV where the impactors in 2182 are
located. The spacing between consecutive VAs corresponds to
$\Delta\sigma=0.0025$.}
\label{fig:omega2162}
\end{figure}

Phase 4 (2162--2182) begins with the shallow 
(for the LOV portion we are discussing)
approach near the
ascending node in 2162, and leads to the deep encounter near the
ascending node in 2182 with a reduced stretching and two separate 
VIs. This phase has to be investigated by using the theory of
\emph{resonant returns} \cite{an10}, \cite{analytic}, 
\cite{clomon2}, see below.
  
The sequence of phases 1, 2 and 3 is not that unusual. For (99942)
Apophis, phase 1 ---in which the Yarkovsky effect is the dominant
source of uncertainty \cite{chesleyACM05, giorgini08} ---
lasts until the 2029 very close approach to the Earth. Phase 2
contains only the 2029 encounter, which increases the spread in $a$,
resulting in an increase of the stretching at the next encounter by a
factor $\simeq 40,000$. In phase 3, leading to a possible impact in
2036, the effect of Yarkovsky and other nongravitational perturbations
is not important in determining the outcome.

Phase 4 in the evolution of our 2182 VIs is a
manifestation of an \emph{interrupted resonant return}. As shown in
Fig.~\ref{fig:bpl2182}, the line of VAs arriving on
the 2182 target plane has a fold, with the tip near the
Earth impact cross section. In the cases of interrupted resonant
returns described so far \cite{pssimp} the fold occurs
before the Earth is reached while, in 2182, 1999 RQ$_{36}$ can
have what we may call a \emph{failed interrupted resonant return}: the
line of VAs turns back on the TP too late, beyond the Earth cross
section.

This fold ---or, better, the entire $\Omega$-shaped set of four folds
shown in Fig.~\ref{fig:omega2162}--- was generated in 2162. For
example, if we use the CLOMON2 run with $da/dt=-17.1\times 10^{-4}$
AU/My, the closest approach in 2162, at $14.2$ Earth radii, occurs for
a value of the LOV parameter\footnote{Note that the
$\sigma$ used here to parameterize the LOV is from CLOMON2 runs for
fixed $da/dt$, and thus is not the same as that of
Table~\ref{t:rq36_risk}; the values we mention are from the run with
$da/dt=-17.1\times10^{-4}\; \mathrm{AU/Myr}$.}  $\sigma=+0.868$,
while the two 2182 VIs are centered at $\sigma=-0.267$ and
$\sigma=-0.361$.
However, one of the folds generated by the 2162 close approach travels
along the LOV towards negative values of $\sigma$: in 2170 it has
reached $\sigma=0.2$, and by the time of the 2182 close approach the
fold is at $\sigma=-0.311$, right in between the two VIs.
 
Figure~\ref{fig:omega2162} shows the $\Omega$-shaped perturbation
introduced in the LOV by the 2162 close approach to the Earth. As
expected from the analytic theory \cite{analytic}, \cite[Section
3.2]{clomon2}, a planet passing near the sequence of VAs introduces
four folds and opens a gap. The two folds corresponding to the
stronger perturbations have the special property that the decrease in
stretching is roughly compensated by increase in width 
\cite[Appendix B]{clomon2}, while the two corresponding to weaker
perturbations result in a smaller increase in stretching without a
corresponding decrease of width.

In 2163, the larger VIs of 2182 are \emph{near} one of the folds of
the latter kind, but not \emph{at} the fold, as shown in
Fig.~\ref{fig:omega2162}. The fold moves along the LOV like a wave
because there is a gradient of semimajor axis ---along the LOV segment
that we are considering--- that has the same sign as the gradient of
the mean anomaly. Consequently, as time passes, the leading VAs, which
are characterized by a smaller mean motion, start to lag, so that the
$\sigma$-coordinate of the fold changes, and the fold point rolls
along the LOV. In 2182 it has moved so much that it happens to be just
between the two strings of consecutive VIs shown in
Figs.~\ref{fig:after80VIs} and \ref{fig:bpl2182}.  The close packing
of VAs near the tip of the fold reflects the drop of the stretching
$S$ in phase 4, as shown in Fig.~\ref{fig:lyap}.

\section{What happens next?}
\label{s:next}

Given a possibility of impact so remote in the future, the question
arises whether there can be changes in the situation in the short
term. 1999 RQ$_{36}$ is currently behind the Sun and cannot be
observed until the spring of 2011. In that year it will undergo a
close approach to Earth with a minimum distance $0.177$ AU on
September 11, 2011.

\begin{figure}[t!]
\centerline{\includegraphics[width=10cm]{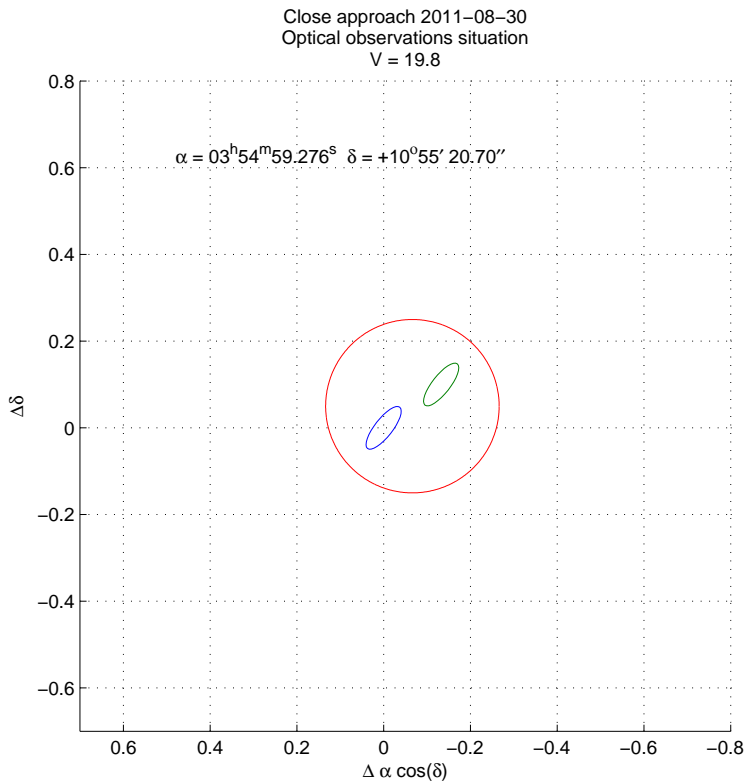}}
\caption{Celestial sphere coordinates (right ascension and
declination, in arcsec) relative to the nominal prediction without
Yarkovsky for the recovery of 1999 RQ$_{36}$ on August 30.0, 2011. The
confidence ellipses are drawn for both cases: no-Yarkovsky (on the
left) and $da/dt=-17.1\times 10^{-4}$ AU/My (on the right). The circle
has a radius of $0.2$ arcsec.}
\label{fig:rec2011}
\end{figure}

The most favorable time for optical observations will be around August
30, with a formal prediction uncertainty of $0.02$ arcsec (longest
axis of $\sigma=1$ confidence ellipse). Figure~\ref{fig:rec2011} shows
the confidence ellipse computed from an orbital solution with no
Yarkovsky effect (on the left) and the one computed from an orbital
solution with a value of the secular $da/dt=-17.1\times 10^{-4}$
AU/My (on the right). The two confidence ellipses, drawn for a
$3\sigma$ confidence level, are indeed disjoint, which means that
there is a significant signal to discriminate between the two
cases. The problem is that the noise contained in the 2011 optical
observations is likely to confuse the results. We have drawn in the
figures a circle with a radius of $0.2$ arcsec, which is larger than
the difference in the predictions due to the Yarkovsky effect to be
measured. Note that by taking many accurate observations it would
certainly be possible to average out the random astrometric errors at
a level lower than the Yarkovsky signal, but this would not at all
solve the problem, because systematic errors in the astrometric data
would not be removed.

In fact, as pointed out by \cite{carpino}, strong correlations and
biases in the astrometric errors, presumably due mostly to catalog
regional errors \cite{tholenDPS,silvaneto}, at present do not allow
absolute astrometry better than about $0.2$ arcsec. It is possible
that by 2011 this problem will be solved with the production
of a statistically sound astrometric error model, removing the main
biases and accounting correctly for the correlations; work to this
goal is currently under way \cite{baerDPS}.

\begin{figure}[t]
\centerline{\includegraphics[width=10cm]{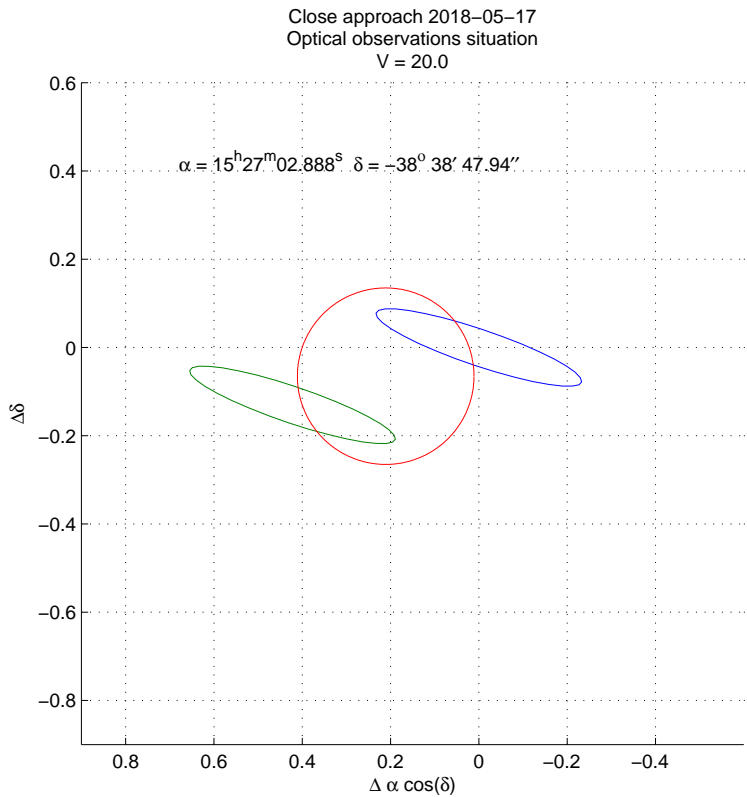}}
\caption{Prediction for recovery of 1999 RQ$_{36}$ on May 17.0, 2018,
plotted with confidence ellipses as in the previous figure: the
no-Yarkovsky case (on the right) and $da/dt=-17.1\times 10^{-4}$ AU/My
(on the left). The circle has a radius of $0.2$ arcsec.}
\label{fig:rec2018}
\end{figure}

In May 2018 there are again favorable conditions to observe 1999 RQ$_{36}$
and to obtain optical astrometry. As shown in Fig.~\ref{fig:rec2018},
at that time the signal caused by the Yarkovsky effect would be
marginally significant with respect to the level of error
corresponding to the current state of the art. However, by that time
it can be presumed that the quality control of optical astrometry
would have progressed, both because of a better statistical quality
control on the past observations and because of the results of the
next generation surveys, such as Pan-STARRS and LSST. In 2024 the
situation would be even better, with a Yarkovsky signal of the order
of 1 arcsec which should provide a tight constraint on the value of
$da/dt$.

To acquire information on the Yarkovsky effect acting on 1999
RQ$_{36}$ in a shorter time frame the best opportunity is to use radar
astrometry during a comparatively close approach in 2011. With a
distance exceeding 26 million km this requires a very powerful radar
system, but the declination at the time of closest approach is
$+20^\circ$, thus the observing conditions are favorable for the
Arecibo radar, which has observed (99942) Apophis at a larger
distance. We have assumed that the measurement error would be similar
to the one of the Arecibo observations of Apophis in January 2005,
that is RMS in range of $0.6$ km and in range-rate of $1$ km/d, and
simulated\footnote{We have used in the simulation observations from
the Arecibo planetary radar. The distance of 1999 RQ$_{36}$ during the
2011 close approach and the size of the object would make an
observation from Goldstone difficult.} two range and two range-rate
observations on 29-31 August 2011. With these prospective
measurements, we have repeated the procedure of Section~\ref{s:yarko}
and obtained an estimated $da/dt$ uncertainty of $\pm
0.25 \times 10^{-4}\; \mathrm{AU/Myr}$. Thus the 2011
observations from Arecibo should reduce the RMS uncertainty by a
factor $\simeq 40$ with respect to the analogous result
with the present data, and by a factor $\simeq 20$ with
respect to the a priori uncertainty in the model of
Eq.~(\ref{eq:dadt}).

Of course, the most likely outcome would be a contradiction between
the radar data and the large 2182 VIs. However, judging from
Fig.~\ref{fig:after80_adadt}, it is very likely there would be other
VIs which would be confirmed and could increase their estimated
IP. For any VIs that remain near nominal, the increase of the
probability density due to the decreased uncertainty could result in
an increase of the IP by a factor $\simeq 15$. Anyway there would be a
very significant increase in the level of information on this orbit,
resulting in changes to the VIs and their IPs.

There is another way to improve the knowledge of the nongravitational
perturbations acting on 1999 RQ$_{36}$: by close-up observations from
a spacecraft it is possible to determine accurately the parameters of
Table~\ref{t:poles}. With an orbiter it would be possible both to
develop a full nongravitational perturbations model, by using visible
and infrared images, and to measure the corresponding perturbations
with a radioscience experiment. 1999 RQ$_{36}$ was the target of the
mission OSIRIS proposed to NASA, and is among the possible targets of
the Marco Polo mission proposed to ESA.\footnote{Neither of the two
missions is presently fully approved and funded.} Improving the
fidelity of asteroid thermal measurements and models, which will
allow, among other applications, the development of more accurate
models for the Yarkovsky effect, should be included among the main
scientific goals of the next generation of asteroid missions.

\section{The deflection problem}
\label{s:deflection}

To complete the worst case analysis, we need to discuss how difficult
it would be to deflect 1999 RQ$_{36}$ from a collision course in case it was
confirmed by very accurate observations in the next decades. The
answer is implicitly contained in Fig.~\ref{fig:lyap}: a deflection
maneuver performed before ``phase 2'', the sequence of close approaches of
2060--2080, is comparatively easy, after 2080 the challenge increases dramatically. 

On the target plane of the 2060 encounter the \emph{keyhole}
\cite{analytic,chodas99}, that is the pre-image of the Earth in the
2182 TP, for a fixed value of $da/dt$ and for a given VI, is a thin
crescent with width less than a km. The TP trace of the VI is the
intersection of the TP confidence region with this crescent: a
deflection can be obtained by achieving a displacement in 2060 by an
amount of this order.

Thus, if the deflection maneuver was performed after the Yarkovsky
effect has been well constrained but before the beginning of phase 2,
that is before 2060, the required change in velocity of the asteroid
would be very small, to the point that it would not be a problem even
by the technology available today, e.g., with the kinetic deflection
method\footnote{For a description of this method, see the ESA study
Don Quijote \cite[Chap. 14]{milanigronchi09}.}.
\begin{figure}[h]
\centerline{\includegraphics[width=10cm]{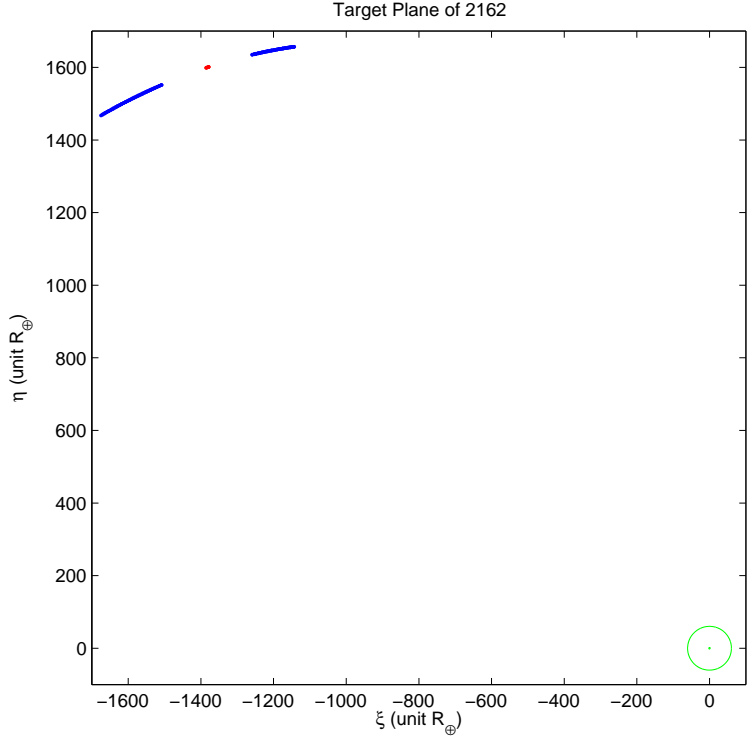}}
\caption{The double keyhole for the impact in 2182 as seen on the
target plane of the 2162 encounter. We have plotted also the points
corresponding to the fold, that is corresponding to the lower tip just
below the Earth impact cross section in Fig.~\ref{fig:bpl2182}. For
comparison, we have plotted a circle around the Earth with radius
equal to the mean radius of the Moon's orbit.}
\label{fig:keyhole_2162}
\end{figure}

Immediately after the 2080 encounter, the amount of velocity change
needed to deflect away from the 2182 impact would be much larger: as
shown in Figure~\ref{fig:after80VIs}, a change in semimajor axis to
deflect out of each of the two largest VIs would be of the order of
several tens of km.

In phase 3 the stretching increases slowly, and so does the required
velocity change, with a peak value in 2162. On the target plane of the
2162 encounter, the two VI traces of 2182 appear as two very long
stretches with length $\simeq 187$ and $\simeq 118$ Earth radii,
respectively, as shown in Fig.~\ref{fig:keyhole_2162}.  Because of the
\emph{failed interrupted resonant return}, the stretching between 2162
and 2182 actually decreases, thus ``the keyhole is larger than the
door'', that is the TP points leading to the impact form a figure with
a width comparable to the diameter of the Earth impact cross session
and a much larger length.  To deflect with a maneuver shortly before
2162 would need an off-track impulse, changing the 2162 TP trace by a
significant fraction of $R_\oplus$.

The current impact monitoring covers a time span of about a
century. If this were to continue, and the asteroid 1999 RQ$_{36}$
were indeed on a collision course for 2182, then the warning about this
would be issued only in 2082, that is at a time when the opportunity
to deflect with requirements compatible with current technology would
already be expired.  To extend the predictability horizon of impact
monitoring seems to us to be a better solution, at least economically,
than waiting for future and hypothetical technological advances.

\section{Conclusions and Future Work}
\label{s:future}

The analysis presented in this paper represents a step toward the
extension of the predictability horizon for Earth impacts of NEAs to a
time scale of several centuries.  The main reason for trying to push
this horizon further away in time is that, in case a deflection is
needed for an Earth impactor, the available time is a non-renewable
``scarce resource''. The example we have analysed shows that the cost
of deflection, although it is generally growing when waiting before 
the action, may remain at very high levels for extended periods of time 
before the impact date. 

Of course, longer time spans imply more room for the NEA to display
the chaoticity of its motion, especially in cases, like the one
discussed here, of rather frequent planetary encounters.
Concerning the understanding of encounter-dominated, strongly chaotic
orbits, the analytical theory developed so far \cite{analytic,clomon2}
has been strongly confirmed, but it does not yet address the issue of
how folds, once developed at an encounter, propagate along the LOV.
This aspect is now the object of ongoing research by our team.

The dynamics over longer times of moderate size asteroids is also
affected by nongravitational forces.  As a consequence, the effort
must pursue significant improvements of the modeling of the various
nongravitational forces that can affect NEAs on multi-century time
scales.
Such progress depends also upon theoretical studies, but mostly upon
the availability of stringent observational constraints, which can be
provided by increased accuracy optical astrometry (with reliable error
models), radar astrometry, physical characterization observations from
the ground and from space, including in situ missions. 

The results on the specific case of the asteroid 1999 RQ$_{36}$ have
been obtained with a dedicated effort, and at the moment we do not
have the capability to extend a comparable scan for impacts over a
longer time span and over a wider range of perturbation parameters to
all the objects with good enough orbits.  Whether the results obtained
for 1999 RQ$_{36}$ are typical rather than exceptional we do
not know yet.  

However, we already know another example of a numbered NEA for which
impact monitoring can be extended beyond the 100 years horizon:
(153814) 2001 WN$_5$. In comparison with 1999 RQ$_{36}$ this is a
somewhat larger asteroid\footnote{2001 WN$_5$ has absolute magnitude
$H=18.2$ instead of $20.8$ for 1999 RQ$_{36}$. However, the size is
very poorly constrained because the taxonomic type is still unknown.},
with no radar observations and therefore a less well determined orbit:
RMS$(a)\simeq 15$ km. In 2028 this asteroid has a deep encounter
(minimum distance $\simeq 40\; R_\oplus$), resulting in a sharp
increase of the stretching, which we can describe as the beginning of
phase 2.  

The main difference with the main case discussed in this paper is that
for 2001 WN$_5$ the Yarkovsky effect does not matter in the impact
monitoring, even for an extended time span, because the effect of the
possible values of $da/dt$ over the timespan of phase 1, that is until
2028, is much smaller than RMS$(a)$.  Given the possible size of this
object, a plausible range of secular semimajor axis drift due to
Yarkovsky could be in the range $|da/dt| < 12 \times 10^{-4}$ AU/My;
the sign is not known because we have no information on the spin axis.
This translates in changes of $a$ in 20 years up to $3.6$ km,
corresponding to about $1/4$ of RMS$(a)$: the large
relative uncertainty on the Yarkovsky effect is not the main
contributor in the uncertainty budget for the 2028 TP trace. Thus the
trace of the confidence region on the 2028 TP is only slightly larger
when considering the uncertainty in $da/dt$ with respect to the one
from a purely gravitational model.

\begin{figure}[h]
\centerline{\includegraphics[width=10cm]{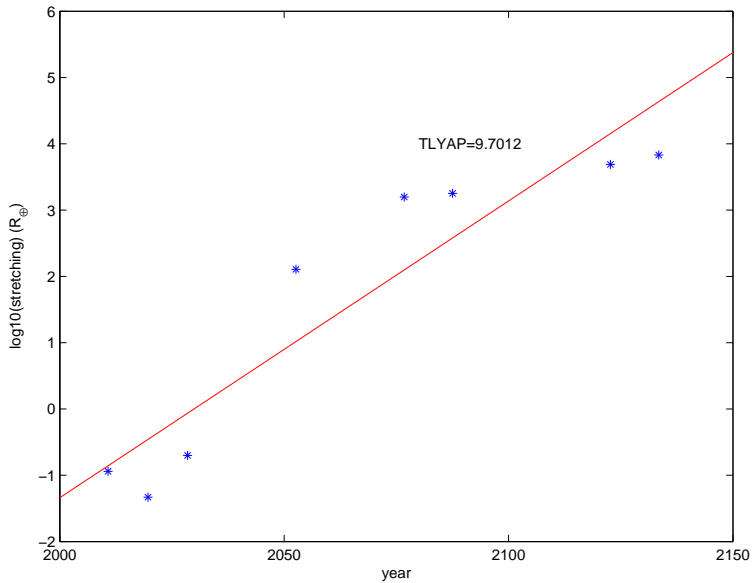}}
\caption{The stretching (in a $\log_{10}$ scale) at each close
  approach with Earth for the sample orbit of 2001 WN$_5$ nearest to
  the center of the 2133 VI. The stretching on average grows by a
  factor $\exp(1)$ on average every $9.7$ years.}
\label{fig:153814lyap}
\end{figure}

This implies that it is possible to use the currently operational
impact monitoring systems CLOMON2 and SENTRY, just extending the time
horizon, to find VIs and estimate the corresponding IP. The result has
been to find a comparatively large VI for the year 2133, with an
estimated IP of $10^{-5}$. Figure~\ref{fig:153814lyap} shows the
behavior of the stretching $S$, showing clearly a phase 1 with modest
growth of $S$ until 2028, a phase 2 containing the 2028 encounter only
(very much like Apophis), followed again by a moderate growth of $S$
in phase 3. The value of the IP for the 2133 VI is not as big as that
for 1999 RQ$_{36}$ in 2182 because there is no phase 4, that is no
interrupted return. Still, this could become a critical case for deflection 
if the VI of 2133 were confirmed by post-2028 observations, because
the amount of required deflection would already be large.

The conclusion we can draw is that the set of NEAs with low MOID and
well determined orbits could contain a significant number of cases for
which an extended impact monitoring would give interesting results,
but we should not expect that these additional cases would have the
same properties of the ones already found.

Another implication is that impact monitoring needs to be continued
forever: even when an asteroid has an extremely well determined orbit,
at some future time it could undergo a ``phase 2'' with one or more
very close approaches, increasing the stretching to the point that
collision solutions become possible later, and would become easily
detectable from the orbit solution with observations during ``phase
3''. This is already known, for a very far future, for (4179) Toutatis
\cite{ostro99} and (29075) 1950 DA \cite{1950DA}.

To be ready to extend the predictability horizon for impact
monitoring, and to handle the unknown parameters contained in the
nongravitational perturbations, we need to set up our algorithms for
full orbit determination in a space of dimension at least 7 (including
either $da/dt$ or similar empirical parameters), for the computation
of the LOV in such space, and for minimization of the close approach
distance in this context. Moreover, the numerical accuracy
requirements are severe and need to be carefully addressed for
reliable semiautomatic operations.

\section*{Acknowledgments}

We wish to dedicate this paper to the memory of Steve Ostro
(1946-2008), who pioneered the radar observations of asteroids, so
essential for these studies.

We thank Mike Nolan and collaborators for making available their
preliminary analysis of the spin axis orientation of 1999 RQ$_{36}$,
and Carl Hergenrother for providing a precise estimate of the spin
rate of the asteroid.

The authors have been supported by: the Italian Space Agency, under
contract ASI/INAF I/015/07/0, Tasks 3130 and 3430 (A.M.,
G.B.V. and F.B.); the Spanish {\em Ministerio de Ciencia y
Tecnolog\'{\i}a} through the grant \mbox{AYA2007-64592} and by the
{\em Junta de Castilla y Le\'on} through the grant \mbox{VA060A07}
(M.E.S. and O.A.). The work of S.R.C. was conducted at the Jet
Propulsion Laboratory, California Institute of Technology, under a
contract with the National Aeronautics and Space Administration.

\end{document}